# EVALUATION OF PHENOMENOLOGICAL CHARACTERISTICS OF THE TWO-PROTON DECAY OF THE $^{45}$FE NUCLEUS

D. E. Lyubashevsky[1], S. G. Kadmensky[1], J. D. Shcherbina[1]

[1]*Voronezh State University, Voronezh, 394018, Russia*

A Green-function formalism describing sequential and virtual two-proton (2p) radioactivity is developed on the basis of the multiparticle theory of one-proton decay. In the proposed approach, two-proton emission is treated as two consecutive one-proton transitions connected by the Green's function of the intermediate nucleus, allowing both on-shell and off-shell intermediate states to be consistently included. Analytical expressions are derived for the total and partial 2p-decay widths as well as for the angular distribution of the emitted protons.

The formalism is applied to the ground-state two-proton decay of $^{45}$Fe leading to the ground state of $^{43}$Cr within the superfluid nuclear model. Calculations demonstrate a pronounced dependence of the decay characteristics on the choice of the single-particle nuclear potential. It is shown that an appropriate choice of the shell potential allows the experimental decay width and the measured proton angular distribution to be reproduced simultaneously. These results indicate that the proposed Green-function approach provides a consistent description of virtual sequential two-proton decay and offers an alternative framework for the interpretation of true two-proton radioactivity.

## INTRODUCTION

Two-proton (2p) radioactivity is one of the most intriguing decay modes of nuclei located beyond the proton drip line. Unlike one-proton (1p) emission, the 2p-decay process is governed not only by the properties of individual proton emission but also by correlations between the emitted protons and the structure of intermediate nuclear configurations. Consequently, the development of theoretical approaches capable of consistently describing different mechanisms of two-proton emission remains an important problem of modern nuclear physics.

The theoretical description of 1p radioactivity has reached a high level of maturity. The characteristics of 1p decays from ground and isomeric states of both spherical and deformed nuclei have been investigated experimentally in detail (see the review in Ref. [1]) and are successfully described within the framework of the multiparticle theory developed on the basis of the approaches originally proposed for α decay [2,3]. This formalism has subsequently been extended to diagonal and off-diagonal one-proton decays [4–7] and employs integral expressions for calculating decay widths.

Following the prediction of two-proton radioactivity by Goldansky [8], extensive experimental and theoretical studies of this decay mode have been carried out (see the review in Ref. [9]). Depending on the energetics of the intermediate nucleus, existing theoretical approaches distinguish two principal decay mechanisms: sequential and true two-proton emission.

Sequential two-proton decay corresponds to the emission of two protons in two successive one-proton transitions through real intermediate nuclear states, when both one-proton decay energies are positive.

Under these conditions, the interaction between the emitted protons can be neglected to a good approximation, and the decay can be described as two independent one-proton processes. Such an approach forms the basis of the R-matrix description of nuclear reactions involving unstable intermediate states [11] and of the kinetic-equation formalism for successive radioactive decays [12]. Sequential two-step decays of spherical nuclei were investigated in detail in Ref. [10].

In contrast, true two-proton radioactivity occurs when at least one of the one-proton decay channels is energetically closed, whereas the total two-proton decay energy remains positive. In this case, the decay proceeds through genuine three-body dynamics, making it impossible to represent the total decay width as a product of two independent one-proton amplitudes. This problem has been successfully addressed within the hyperspherical three-body formalism developed in Refs. [13–15] and reviewed in Ref. [9], where both nuclear and Coulomb interactions between the three fragments are treated explicitly.

The first experimental observation of true two-proton radioactivity was reported for the ground-state decay of the **$^{45}$Fe** nucleus [17,18], confirming the earlier prediction of Ref. [16]. Since then, this nucleus has become one of the benchmark systems for investigations of two-proton emission. In addition to the total decay width [17,18], measurements of the energy [19] and angular [20,21] distributions of the emitted protons have provided stringent tests for theoretical models. While the diproton model based on the virtual $^{2}$He state [22–25] was unable to reproduce the experimental observations satisfactorily, significantly better agreement was obtained within the three-body R-matrix approach [26]. Nevertheless, this description employs effective three-body interactions whose relation to the shell-model potentials and superfluid nucleon-nucleon interactions commonly used in nuclear-structure calculations remains an open question [26].

An alternative description of two-step two-proton decay was proposed in Refs. [27–29]. In that approach, the decay is represented as two successive one-proton emissions connected through the intermediate nucleus, and the corresponding decay widths are constructed from integral expressions for one-proton decay amplitudes [30–32]. The formalism employs conventional shell-model potentials together with the superfluid description of nucleon pairing. However, the previous formulation was restricted to the case of delayed sequential proton emission and did not explicitly include contributions associated with virtual intermediate configurations.

Among the continuum shell-model formulations, a fully microscopic treatment of two-proton emission was developed by Rotureau et al. [33] within the Shell Model Embedded in the Continuum (SMEC) framework. By explicitly partitioning the total Hilbert space into bound discrete configurations ($\mathcal{Q}$), one-proton continuum ($\mathcal{P}$), and two-proton continuum ($\mathcal{T}$), SMEC provides a rigorous unified description of nuclear structure and reaction dynamics. Depending on the relative energy and coupling matrix elements between these subspaces, SMEC distinguishes three asymptotic modes of two-proton emission:

1. Direct emission via three-body asymptotics (or diproton cluster decay): Arising from direct transitions from the bound $\mathcal{Q}$ subspace to the two-particle continuum $\mathcal{T}$ via $H_{\mathcal{Q}\mathcal{T}}G_{\mathcal{T}}^{(+)}H_{\mathcal{T}\mathcal{Q}}$, effective when intermediate $A-1$states are energetically closed ($\mathcal{P}$-subspace coupling is suppressed).
2. Sequential decay through intermediate resonances: Proceeding via real intermediate resonance states in the $A-1$ nucleus embedded in the $\mathcal{P}$ subspace.
3. Virtual sequential decay through the correlated continuum: Proceeding through the off-shell continuum of the $A-1$ system in the absence of real, energetically accessible intermediate resonances.

While SMEC offers a comprehensive microscopic treatment including full configuration mixing and channel antisymmetrization, its practical implementation requires extensive multi-configuration shell-model calculations and effective zero-range residual interactions. In contrast, the Green-function formalism proposed in the present work constructs two-proton decay width directly from single-proton transition potentials and superfluid nucleon-nucleon pairing correlations, providing a computationally transparent alternative to describe both real on-shell and virtual off-shell intermediate transitions.

In the present work, we extend this approach by developing a general Green-function formalism for diagonal two-step two-proton decays of spherical and deformed nuclei in both ground and excited states. The proposed formulation consistently incorporates both on-shell and off-shell intermediate nuclear states through the Green's function of the intermediate nucleus. As a result, the total decay width is naturally separated into sequential and virtual contributions, providing a unified description of real and virtual two-step proton emission within the same theoretical framework. The formalism is applied to the ground-state two-proton decay of $^{45}$Fe leading to the ground state of $^{43}$Cr, where the total decay width and proton angular distribution are calculated within the superfluid model of the atomic nucleus [34].

# I. FORMALISM OF SINGLE-PROTON DECAY

To formulate a general description of two-proton decay, we first briefly summarize the theoretical formalism of single-proton (1p) decay developed in Refs. [2–7,10], which constitutes the basis of the present approach. We consider the decay of a spherical or deformed parent nucleus $A$ from an arbitrary ground or excited state $i$, characterized by the center-of-mass momentum, total angular momentum, its projection, and additional quantum numbers. The wave function of the parent nucleus in the state $i$ is written as

$$\psi_i = \frac{1}{\sqrt{L^3}} e^{\frac{i(\mathbf{P}_A, R_A)}{\hbar}} \psi_{\sigma_A}^{J_A m_A} \tag{1}$$

where the first factor is the plane wave describing the center-of-mass motion of the parent nucleus, normalized to unity within a cubic volume with periodic boundary conditions, whereas the second factor represents the intrinsic wave function of the nucleus with internal energy. The latter is related to the binding energy of the parent nucleus by

$$E(Z,\ A,\ J_A,\ \sigma_A) = E_0(Z,\ A) + U(Z,\ A,\ J_A,\ \sigma_A), \tag{2}$$

where $E_0(Z,A)$ is the ground-state energy of the nucleus and $U\ (Z,A,J_A,\sigma_A)$ is its excitation energy. Accordingly, the total energy $E_i$ of the parent nucleus is given by

$$E_i = E(Z,A,J_A,\sigma_A) + \frac{P_A^2}{2M_A}. \tag{3}$$

At 1p decays the nucleus $(Z,A)$ passes from the initial state $i$ to the final state $f = p_1\ m_1\ f_1$, corresponding to the ejected proton with momentum $p_1$ and its spin projection $m_1$ and to the state $f_1$ of the daughter nucleus $(Z-1, A-1)$ with wave function $\psi_{f_1}$ and energy $E_{f_1}$ defined by formulas (1) and (3) when index $i$ is replaced by index $f_1 = \mathbf{P}_{A\text{-}1}\ J_{A\text{-}1}\ m_{A\text{-}1}\ \sigma_{A\text{-}1}$, including momentum $\mathbf{P}_{A\text{-}1}$, center of mass of the daughter nucleus with coordinate $\mathbf{R}_{A\text{-}1}$ and internal state of the daughter nucleus with wave function $\psi_{\sigma_{A-1}}^{J_{A-1}m_{A-1}}$ Introducing coordinates $\mathbf{R}_A$, $\mathbf{r}$ and impulses $\mathbf{P}_f$, $\mathbf{p}$, related to the initial coordinates $R_{A-1}$, $\mathbf{r}_1$ and impulses $\mathbf{P}_{A-1}$, $\mathbf{p}_1$ by the relations:

$$\begin{cases} r = \mathbf{r}_1 - R_{A-1}\ , \\ \mathbf{R}_A = \dfrac{M_{A-1}R_{A-1} + m_p\,\mathbf{r}_1}{M_A}, \end{cases} \qquad \begin{cases} \mathbf{P}_f = \mathbf{p}_1 + \mathbf{P}_{A-1}\ , \\ p = \dfrac{M_{A-1}\mathbf{p}_1 - M_p\mathbf{P}_{A-1}}{M_A}, \end{cases} \tag{4}$$

where $M_p$ , $M_A$ , $M_{A-1}$ denote the masses of the proton, parent nucleus, and daughter nucleus, respectively, the wave function $\psi_f$ of the final state can be written, following Refs. [2–7,10], in the form

$$\psi_f = \frac{1}{\sqrt{L^3}} e^{\frac{i(\mathbf{P}_f\ ,R_A)}{\hbar}} \psi_{\sigma_{A-1}}^{J_{A-1}m_{A-1}} \varphi_{p_1}, \tag{5}$$

The proton wave function entering Eq. (5) is represented as

$$\varphi_{p_1} = \chi_{\frac{1}{2}m_1} \frac{1}{\sqrt{L^3}} \chi_k(\mathbf{r}), \tag{6}$$

where $\chi_{\frac{1}{2}m_1}$ is the proton spin function and $\chi_k(\mathbf{r})$ is the distorted wave describing the relative motion of the emitted proton and the daughter nucleus in the effective interaction potential. For diagonal 1p decays of spherical and deformed nuclei, this potential is determined by the spherical component of the Coulomb interaction between the proton and the daughter nucleus [2–5,7]. For off-diagonal 1p decays of spherical nuclei, the effective interaction is described by the optical potential introduced in Ref. [6], which accounts for both nuclear and Coulomb interactions.

In the following, we restrict ourselves to diagonal 1p decays of spherical and deformed nuclei and to off-diagonal decays of spherical nuclei only. Under these assumptions, the effective interaction becomes spherically symmetric, and the distorted wave can be expanded as

$$\chi_{\mathbf{k}}(\mathbf{r}) = 4\pi \sum_{lm_l} i^l \frac{\varphi_l(k,r)}{kr} Y_{lm_l}(\Omega_{\mathbf{r}}) Y^*_{lm_l}(\Omega_{\mathbf{k}}), \tag{7}$$

where the spherical Bessel functions $j_l(k,r)$ are replaced by the regular radial solutions $\frac{\varphi_l(k,r)}{kr}$ of the Schrödinger equation in the effective potential, satisfying the appropriate boundary conditions and containing the scattering phase shifts.

The total energy $E_f$ of the final state is expressed as

$$E_f = E(Z-1, A-1, J_{A-1}, \sigma_{A-1}) + \frac{\mathbf{p}_1^2}{2m_p} + \frac{\mathbf{P}_{A-1}^2}{2M_{A-1}}, \tag{8}$$

Using the conservation of total energy together with Eqs. (3) and (8), one obtains

$$Q_1\left(J_A,\sigma_A;J_{A-1},\sigma_{A-1}\right) = E\left(Z,\, A,\, J_A,\, \sigma_A\right) - E\left(Z-1,\, A-1,\, J_{A-1},\, \sigma_{A-1}\right), \tag{9}$$

which leads to

$$Q_1\left(J_A,\sigma_A,J_{A-1},\sigma_{A-1}\right)=\frac{\mathbf{p}_1^2}{2M_p}+\frac{\mathbf{P}_{A-1}^2}{2M_{A-1}}-\frac{\mathbf{P}_A^2}{2M_A}, \tag{10}$$

Taking into account the coordinate transformation (4) and the conservation of total momentum,

$$P_A = P_f \tag{11}$$

Eq. (10) can be reduced to the form

$$Q_1\left(J_A,\sigma_A;J_{A-1},\sigma_{A-1}\right)=\frac{\hbar^2 k^2}{2\mu_1}=T \tag{12}$$

where $T$ is the kinetic energy of the relative proton–daughter motion and $\mu_1=\frac{M_p M_{A-1}}{M_A}$ is the corresponding reduced mass.

Within the formalism developed in Refs. [2–7], the total width of the 1p decay of an unpolarized parent nucleus, averaged over the spin projections of the initial state, is given by

$$\Gamma_{p\sigma_A}^{J_A}=\frac{2\pi}{2J_A+1}\sum_{m_A\, f}\left|M_{fi}\right|^2\delta\left(E_f-E_i\right), \tag{13}$$

where the transition matrix element $M_{fi}$ is defined as

$$M_{fi}=\hat{A}\left\{\psi_f \mid \tilde{V}_{\mathrm{p}_1 A-1}\right\}\mid\psi_i, \tag{14}$$

with

(15)

$$\tilde{V}_{p_1 A-1} = \tilde{V}^{d}_{p_1 A-1}(\mathbf{r}) \equiv ReV^{opt}_{p_1 A-1}(\mathbf{r}) - \left(V^{coul}_{p_1 A-1}(\mathrm{r})\right)^0 ,$$

For diagonal 1p decays of spherical and deformed nuclei, the transition potential is determined by Eq. (15) [2–5,7]. In the case of off-diagonal decays of spherical nuclei, it is expressed through the effective proton–nucleon interaction, which, within the theory of finite Fermi systems, coincides with the quasiparticle interaction amplitude in the particle–hole channel [6,35].

It is convenient to introduce the channel function $U_c^{J_A m_A}$

$$U_c^{J_A m_A} = \sum_{m_{A-1} m_j m_l m_1} C^{J_A m_A}_{JA-1 j m_{A-1} m_j} C^{j m_j}_{l \frac{1}{2} m_l m_1} Y_{l m_l}(\Omega_{\mathbf{r}}) \chi_{\frac{1}{2} m_1} \psi^{J_{A-1} m_{A-1}}_{\sigma_{A-1}} \tag{16}$$

where the channel quantum numbers specify the angular momentum coupling scheme. Using the properties of the Clebsch–Gordan coefficients, the final-state wave function $\psi^{J_{A-1} m_{A-1}}_{\sigma_{A-1}} \chi_{\frac{1}{2} m_1} \chi_{\mathbf{k}}(\mathbf{r})$ (5) can be expanded as

$$\psi^{J_{A-1} m_{A-1}}_{\sigma_{A-1}} \chi_{\frac{1}{2} m_1} \chi_{\mathbf{k}}(\mathbf{r}) = 4\pi \sum_{l j m_l m_j \, J_A m_A} i^l \frac{\varphi_l(k,r)}{kr} Y^*_{l m_l}(\Omega_{\mathbf{k}}) C^{J_A m_A}_{J_{A-1} j m_{A-1} m_j} C^{j m_j}_{l \frac{1}{2} m_l m_1} U_c^{J_A m_A} . \tag{17}$$

Substituting Eq. (17) into Eq. (14) and integrating over the center-of-mass coordinate, the transition matrix element $M_{fi}$ assumes the form

$$M_{fi} = \frac{4\pi}{\sqrt{L^3}} \delta_{P_A, \mathbf{P}_f} \frac{1}{k} M^{J_A m_A}_{\sigma A}\left(J_{A-1}, m_{A-1}, \sigma_{A-1}, m_1, \mathbf{k}\right), \tag{18}$$

where

$$M^{J_A m_A}_{\sigma_A}\left(J_{A-1}, m_{A-1}, \sigma_{A-1}, m_1, \mathbf{k}\right) = \sum_{l j m_l m_j} i^l Y^*_{l m_l}(\Omega_{\mathbf{k}}) C^{J_A m_A}_{J_{A-1} j m_{A-1} m_j} C^{j m_j}_{l \frac{1}{2} m_l m_1} M^{J_A}_{\sigma_A c}(T), \tag{19}$$

and the reduced matrix element $M^{J_A}_{\sigma_A c}(T)$ is given by

$$M_{\sigma_A c}^{J_A}(T) = \hat{A}\left\{U_c^{J_A m_A} \frac{\varphi_l(k,r)}{r} \mid \tilde{V}_{\mathrm{p}_1 A-1}\right\} \mid \psi_{\sigma_A}^{J_A m_A}. \tag{20}$$

Substituting Eq. (18) into Eq. (13), replacing the summation over the proton momentum by integration, and using the energy conservation relation (12), one arrives at the following expression for the total 1p-decay width:

$$\Gamma_{p\sigma_A}^{J_A} = \frac{2\pi}{(2J_A+1)} \sum_{m_A m_{A-1} m_1 J_{A-1} \sigma_{A-1}} \int \left| \sum_{l j m_l m_j} i^l C_{J_{A-1}\, j m_{A-1} m_j}^{J_A m_A} C_{l \frac{1}{2} m_l m_1}^{j m_j} Y_{l m_l}^{*}(\Omega_{\mathbf{k}}) \tilde{M}_{\sigma_A c}^{J_A}(T) \right|^2 d\Omega_{\mathbf{k}}, \tag{21}$$

where the reduced transition matrix element is determined by Eq. (20), with the regular Coulomb radial function replaced by the energy-normalized scattering solution.

Finally, summation over the angular-momentum projections $m_A$ ,$m_{A-1}$ ,$m_1$ using the orthogonality of the Clebsch–Gordan coefficients yields the isotropic angular distribution of protons emitted from an unpolarized parent nucleus. Accordingly, the total width can be represented as the sum of partial widths $\Gamma_{p\sigma_A c}^{J_A}(T)$,

$$\Gamma_{p\sigma_A}^{J_A} = \sum_c 2\pi \left|\tilde{M}_{\sigma_A c}^{J_A}(T)\right|^2 \int \frac{d\Omega_{\mathbf{k}}}{4\pi} = \sum_c \Gamma_{p\sigma_A c}^{J_A}(T). \tag{22}$$

where the partial proton widths $\Gamma_{p\sigma_A c}^{J_A}(T)$ are determined by the integral expression previously derived in Refs. [2–7]:

$$\Gamma_{p\sigma_A c}^{J}(T) = 2\pi \left|\tilde{M}_{\sigma_A c}^{J_A}(T)\right|^2. \tag{23}$$

## II. FORMALISM OF TWO-STEP TWO-PROTON DECAY

Following the single-proton formalism summarized in Sec. I and developed in Refs. [2–7,10], we now derive a general expression for the width of two-step two-proton (2p) decay. We consider the transition of a spherical or deformed parent nucleus from the initial state (i), described by the wave function $\psi_i$ (1) and energy $E_i$ (3), to the final state $f = \mathbf{p}_1\, m_1\, \mathbf{p}_2\, m_2$ ,characterized by

$$E_f = \frac{\mathbf{p}_1^2}{2m_p} + \frac{\mathbf{p}_2^2}{2m_p} + \frac{\mathbf{P}_{A-2}^2}{2m_{A-2}} + E\left(Z-2, A-2, J_{A-2}, \sigma_{A-2}\right), \tag{24}$$

associated with the emission of the first and second protons with coordinates, momenta, and projections of spins $\mathbf{r}_1, \mathbf{p}_1, m_1$ and $\mathbf{r}_2, \mathbf{p}_2, m_2$, respectively, and the formation of a final nucleus $(Z-2, A-2)$ in a state $f_2$ with wave function $\psi_{f_2}$ and energy $E_{f_2}$, defined by formulas (9) and (10) when the index $i$ is replaced by the index $f_2$. In the following, we restrict ourselves to the cases in which the dominant contribution to the two-step decay arises from diagonal one-proton decays of spherical and deformed nuclei and from one-proton decays involving off-diagonal matrix elements for spherical nuclei. Under these assumptions, the two-step decay width of an unpolarized parent nucleus, averaged over the spin projections of the initial state, can be written in the same form as Eq. (13), with the transition matrix element $M_{fi}$ represented diagrammatically in Fig. 1.

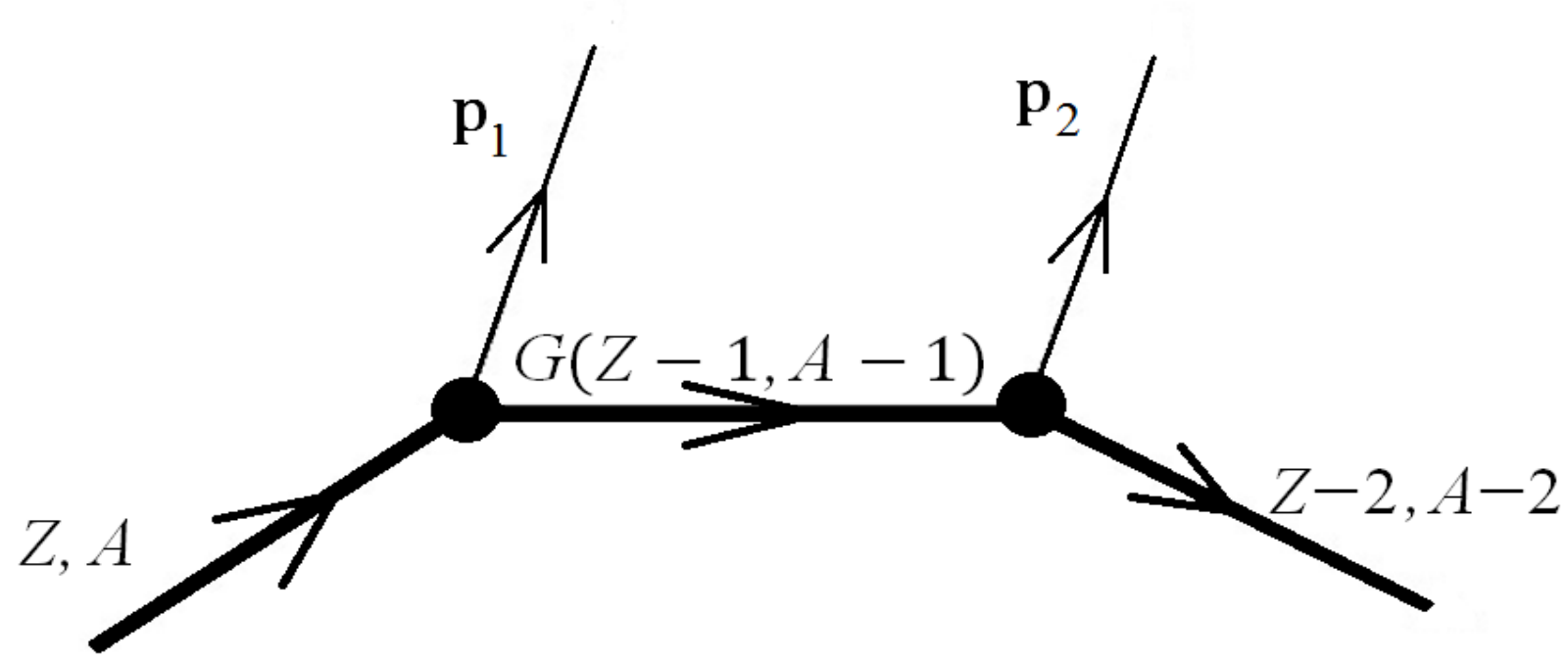

FIG. 1. Diagram illustrating the process of two-step two-proton decay. Here, $(Z, A)$ and $(Z-2, A-2)$ stand for the charges and masses of initial and final nuclei, p1 and p2 are the momenta of emitted protons, and $G(Z-1, A-1)$ is the Green's function for the intermediate nucleus.

In Fig. 1, the filled circles denote the vertex functions describing the interaction of the emitted proton with the residual nucleus through the effective transition potential $\tilde{V}_{\mathrm{p}_1 A-1}$

, whereas the bold directed line represents the one-particle Green's function $G(Z-1,A-1)$ of the intermediate nucleus $(Z-1,A-1)$,

$$G(Z-1,\ A-1)=\sum_{f_1}\frac{\left|\psi_{f_1}\psi_{f_1}\right|}{E_i-E_{f_1}-\dfrac{\mathbf{p}_1^2}{2m_p}+\dfrac{i\,\Gamma_{\sigma_{A-1}}^{J_{A-1}}}{2}},\tag{25}$$

where the finite total width $\Gamma_{\sigma_{A-1}}^{J_{A-1}}$ of the intermediate nuclear state $J_{A-1}\sigma_{A-1}$ is explicitly taken into account. The present formalism assumes that the width of the intermediate state considerably exceeds the total width of the initial parent $(Z-1,A-1)$ state. This condition is naturally fulfilled, for example, for ground-state two-proton decay of even $Z$ nuclei, where proton pairing results in substantially different one-proton decay energies of the parent and intermediate nuclei [10].

To substantiate the assumption that the width of the intermediate state of $\Gamma_{\sigma_{A-1}}^{J_{A-1}}$ significantly greater than the width of the parent nucleus $\Gamma_{\sigma_A}^{J_A}$, let us estimate the value of $\Gamma_{\sigma_{A-1}}^{J_{A-1}}$ for the nucleus $^{44}$Mn within the framework of the same superfluid model. Using formula (23) and the corresponding single-particle amplitudes, we obtain $\Gamma_{\sigma_{A-1}}^{J_{A-1}}\sim 10^{-16}$ – $10^{-15}$ MeV (depending on the potential), which is 2–3 orders of magnitude greater than $\Gamma_{\sigma_A}^{J_A}\sim 10^{-19}$ MeV. Consequently, the condition $\Gamma_{\sigma_{A-1}}^{J_{A-1}} >> \Gamma_{\sigma_A}^{J_A}$ is fulfilled with a large margin, and neglecting the width of the mother state in the denominator of the Green function is correct. Moreover, taking into account the finite width of $\Gamma_{\sigma_A}^{J_A}$ would lead to corrections of the order $\Gamma_{\sigma_A}^{J_A}\big/\Gamma_{\sigma_{A-1}}^{J_{A-1}}\sim 10^{-3}$, which lies far beyond the accuracy of our calculation.

The intermediate state $f_1=\mathbf{p}_{A\text{-}1}\ J_{A\text{-}1}\ m_{A\text{-}1}\ \sigma_{A\text{-}1}$ entering Eq. (25) is described by the wave function and energy obtained from Eqs. (1) and (3) after replacing the index $i$ with $f_1$. Using the energy relation (9), the denominator of the Green's function can be rewritten accordingly.

Depending on its energy, the intermediate nuclear state may correspond either to a real or to a virtual configuration. When the intermediate state satisfies the energy $E_{f_1}$ conservation law $E_{f_1} + \frac{\mathbf{p}_1^2}{2M_p} = E_i$,

$$T = Q_1\left(J_A,\sigma_A;J_{A-1},\sigma_{A-1}\right), \tag{26}$$

it lies on the mass shell of the decaying system and corresponds to a real one-proton decay of the parent nucleus. This situation gives rise to the sequential two-step two-proton decay mechanism. Conversely, if Eq. (26) cannot be fulfilled for any physically accessible energy, the intermediate state remains off shell and contributes only virtually, leading to the virtual two-step two-proton decay mechanism.

To evaluate the corresponding decay width, we introduce the relative coordinates momenta $\mathbf{r},\mathbf{p}$ and $\mathbf{r}',\mathbf{p}'$ of the proton–nucleus subsystems by transformations analogous to Eq. (4). Applying the conservation of energy $E_f = E_i$ together with Eqs. (3) and (24), we obtain

$$Q_0\left(J_A,\sigma_A;J_{A-2},\sigma_{A-2}\right)=\frac{\mathbf{p}^2}{2\mu_1}+\frac{\mathbf{p}'^2}{2\mu_2}=T+T', \tag{27}$$

where the total two-proton decay energy $Q_0\left(J_A,\sigma_A;J_{A-2},\sigma_{A-2}\right)$ is expressed as

$$Q_0\left(J_A,\sigma_A;J_{A-2},\sigma_{A-2}\right) = E\left(Z, A, J_A, \sigma_A\right) - E\left(Z-2, A-2, J_{A-2}, \sigma_{A-2}\right) = Q_1\left(J_A,\sigma_A;J_{A-1},\sigma_{A-1}\right) + Q_2\left(J_{A-1},\sigma_{A-1};J_{A-2},\sigma_{A-2}\right) \tag{28}$$

and the remaining quantities denote the reduced mass $\mu_2$ and kinetic energy $T'$ of the second proton relative to the final nucleus $(Z-2, A-2)$.

Using distorted proton wave functions $\bar{V}_{\mathrm{p}_1 A-1}(\mathbf{r})$ and $\bar{V}_{\mathrm{p}_2 A-2}(\mathbf{r})$ of the form introduced in Eq. (7), together with channel expansions analogous to Eq. (17), replacing the momentum summation by integration and applying the energy conservation relation (27), the general expression for the two-step two-proton decay width corresponding to the diagram shown in Fig. 1 becomes

$$\Gamma^{J_A}_{2p\sigma_A} = \frac{2\pi}{2J_A+1} \sum_{m_A m_1 m_2 m_{A-2} J_{A-2}\sigma_{A-2}} \int_0^{Q_0(J_A,\,\sigma_A;\,J_{A-2},\,\sigma_{A-2})} dT \times$$

$$\times \int \left| \sum_{J_{A-1}\sigma_{A-1}m_{A-1}} \frac{\tilde{M}^{J_A m_A}_{\sigma_A}\left(J_{A-1},m_{A-1},\sigma_{A-1},m_1,\mathbf{k}\right)\tilde{M}^{J_{A-1}m_{A-1}}_{\sigma_{A-1}}\left(J_{A-2},m_{A-2},\sigma_{A-2},m_2,\mathbf{k}'\right)}{Q_1\left(J_A,\sigma_A;J_{A-1},\sigma_{A-1}\right)-T+\frac{i\Gamma^{J_{A-1}}_{\sigma_{A-1}}}{2}} \right|^2 d\Omega_{\mathbf{k}}\, d\Omega_{\mathbf{k}'},$$

(29)

where the wave numbers of the emitted protons are determined from

$$T = \frac{\hbar^2 k^2}{2\mu_1}, \quad T' = \frac{\hbar^2 k'^2}{2\mu_2} = Q_0\left(J_A,\sigma_A;\, J_{A-2},\sigma_{A-2}\right) - T\,, \tag{30}$$

and the matrix elements $\tilde{M}^{J_A m_A}_{\sigma_A}\left(J_{A-1},m_{A-1},\sigma_{A-1},m_1,\mathbf{k}\right)$ and $\tilde{M}^{J_{A-1}m_{A-1}}_{\sigma_{A-1}}\left(J_{A-2},m_{A-2},\sigma_{A-2},m_2,\mathbf{k}'\right)$ are obtained from Eq. (19) after replacing the index $J_{A-1}\sigma_{A-1}lm_l\, jm_j$ by the $J_{A-1}\sigma_{A-1}l_1 m_{l_1}\, j_1 m_{j_1}$ and $J_{A-2}\sigma_{A-2}l_2 m_{l_2}\, j_2 m_{j_2}$, respectively.

Integration of Eq. (29) over the emission directions of both protons, followed by the use of the completeness and orthogonality relations of the Clebsch–Gordan coefficients, yields the total two-proton decay width in the form

$$\Gamma^{J_A}_{2p\sigma_A} = \frac{1}{2\pi} \sum_{JA-1\sigma_{A-1}J_{A-2}\sigma_{A-2}j_1 l_1 j_2 l_2} \times$$

$$\times \int_0^{Q_0(J_A,\,\sigma_A;\,J_{A-2},\,\sigma_{A-2})} dT \frac{\Gamma^{J_A}_{p_1\sigma_A J_{A-1}\sigma_{A-1}j_1 l_1}(T)\,\Gamma^{J_{A-1}}_{p_2\sigma_{A-1}J_{A-2}\sigma_{A-2}j_2 l_2}(Q_0-T)}{\left[\left(Q_1\left(J_A,\sigma_A;J_{A-1},\sigma_{A-1}\right)-T\right)^2+\frac{\left(\Gamma^{J_{A-1}}_{\sigma_{A-1}}\right)^2}{4}\right]}, \tag{31}$$

where the partial one-proton widths are defined by expressions analogous to Eq. (22).

For nuclei heavier than the lightest proton emitters, the total width of the intermediate state is generally much smaller than the corresponding decay energy $|Q_1|$. Under this condition, the integral appearing in Eq. (31),

$$I = \frac{1}{2\pi}\int_0^{Q_0} dT\, \Gamma^{J_A}_{p_1\sigma_A J_{A-1}\sigma_{A-1} j_1 l_1}(T)\, \Gamma^{J_{A-1}}_{p_2\sigma_{A-1} J_{A-2}\sigma_{A-2} j_2 l_2}(Q_0 - T)\left[(Q_1 - T)^2 + \frac{\left(\Gamma^{J_{A-1}}_{\sigma_{A-1}}\right)^2}{4}\right]^{-1} \equiv \tag{32}$$

$$\equiv \frac{1}{2\pi}\int_{C_1} f(T)dT \ ,$$

can be evaluated by analytic continuation into the complex energy plane. The integration contour $C_1$ is closed in the upper half-plane $C_2$ as illustrated in Fig. 2.

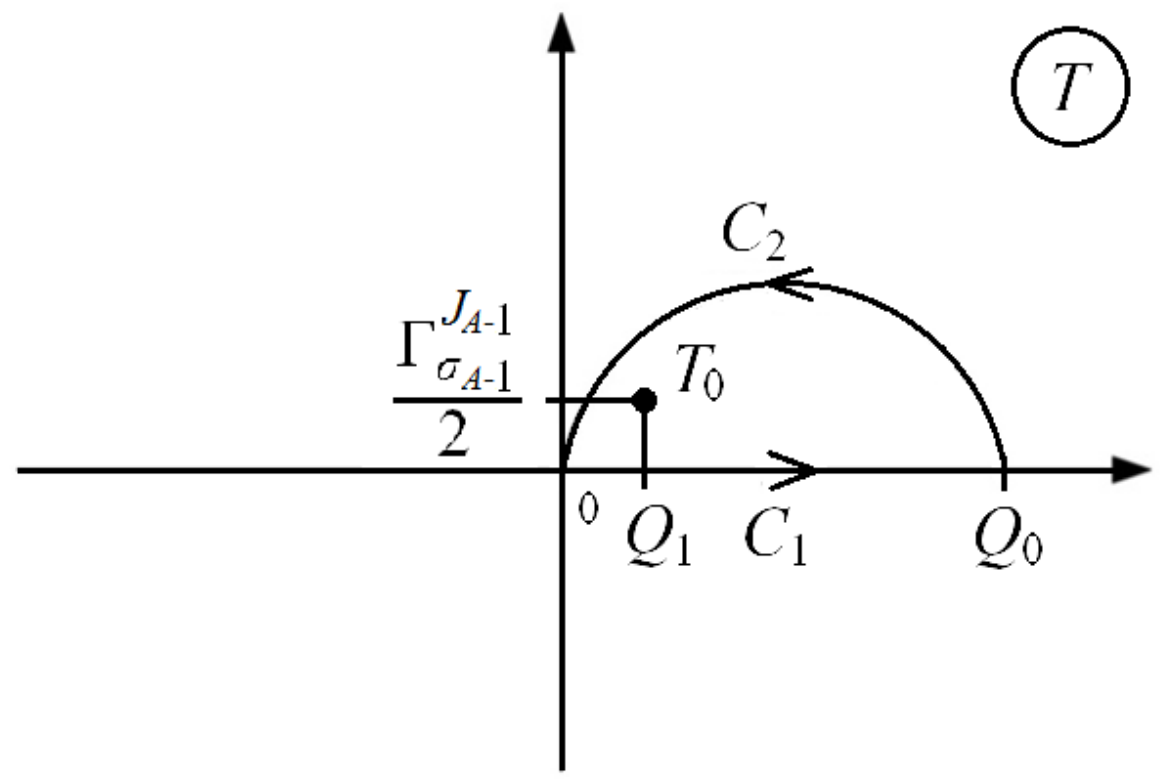

FIG. 2. Integration contour in the complex energy plane. The contour $C_1$ along the real axis is completed by the semicircular contour $C_2$, forming the closed contour $C = C_1 + C_2$. For positive values of the one-proton decay energy $Q_1$, the integrand possesses a pole $T = T_0 = Q_1 + \dfrac{i\,\Gamma^{J_{A-1}}_{\sigma_{A-1}}}{2}$ inside the contour.

The integral $I$ (32) may then be decomposed into two contributions,

$$I = I_1 + I_2 \ , \tag{33}$$

where

$$I_1 = \frac{1}{2\pi}\oint_C f(T)\,dT \ , \qquad I_2 = -\frac{1}{2\pi}\int_{C_2} f(T)\,dT. \tag{34}$$

Application of Cauchy's residue theorem gives

$$I_1 = \frac{\Gamma^{J_A}_{p_1\sigma_A J_{A-1}\sigma_{A-1}j_1l_1}(Q_1)\,\Gamma^{J_{A-1}}_{p_2\sigma_{A-1}J_{A-2}\sigma_{A-2}j_2l_2}(Q_2)}{\Gamma^{J_{A-1}}_{\sigma_{A-1}}} , \tag{35}$$

Substitution of Eq. (35) into Eq. (31) leads to

$$\left(\Gamma^{J_A}_{2p\sigma_A}\right)^{suc} = \sum_{JA-1\sigma_{A-1}J_{A-2}\sigma_{A-2}j_1l_1j_2l_2} \frac{\Gamma^{J_A}_{p_1\sigma_A J_{A-1}\sigma_{A-1}j_1l_1}(Q_1)\,\Gamma^{J_{A-1}}_{p_2\sigma_{A-1}J_{A-2}\sigma_{A-2}j_2l_2}(Q_0-Q_1)}{\Gamma^{J_{A-1}}_{\sigma_{A-1}}} , \tag{36}$$

which represents the width of the sequential two-step two-proton decay associated with two successive real one-proton emissions.

The remaining contribution corresponds to the principal-value part of the integral. When the dominant contribution $|Q_1 - T| \gg \frac{\Gamma^{J_{A-1}}_{\sigma_{A-1}}}{2}$ arises from energies sufficiently far from the pole $T_0 = Q_1 + \frac{i\,\Gamma^{J_{A-1}}_{\sigma_{A-1}}}{2}$, the integral takes the form

$$I_2 = \frac{1}{2\pi}\int_0^{Q_0} dT\Gamma^{J_A}_{p_1\sigma_A J_{A-1}\sigma_{A-1}j_1l_1}(T)\,\Gamma^{J_{A-1}}_{p_2\sigma_{A-1}J_{A-2}\sigma_{A-2}j_2l_2}(Q_0-T)\left[Q_1-T\right]^{-2}. \tag{37}$$

Substituting Eq. (37) into Eq. (31) yields the width of the virtual two-step two-proton decay,

$$\left(\Gamma^{J_A}_{2p\sigma_A}\right)^{v} = \sum_{J_{A-1}\sigma_{A-1}J_{A-2}\sigma_{A-2}j_1l_1j_2l_2} \frac{1}{2\pi}\int_0^{Q_0} dT\Gamma^{J_A}_{p_1\sigma_A J_{A-1}\sigma_{A-1}j_1l_1}(T)\,\Gamma^{J_{A-1}}_{p_2\sigma_{A-1}J_{A-2}\sigma_{A-2}j_2l_2}(Q_0-T)\times$$

$$\times\left[(Q_1-T)\right]^{-2}. \tag{38}$$

Accordingly, when both one-proton decay energies $Q_1$ and $Q_2\,Q_2$ are positive, the total two-proton decay width is given by

$$\Gamma^{J_A}_{2p\sigma_A} = \left(\Gamma^{J_A}_{2p\sigma_A}\right)^{suc} + \left(\Gamma^{J_A}_{2p\sigma_A}\right)^{v}. \tag{39}$$

i.e., as the sum of the sequential and virtual contributions. If one of the one-proton decay energies becomes negative while the total two-proton decay energy remains positive, only the virtual contribution survives.

The expressions obtained for the sequential and virtual decay widths are identical in form to those derived previously for diagonal two-step decays of spherical nuclei in Ref. [10], but are now extended to include diagonal decays of deformed nuclei and transitions involving off-diagonal matrix elements in spherical nuclei. Consequently, the previously established correspondence between the sequential component and the results of the R-matrix theory [11] and the kinetic-equation approach [12] remains valid in the more general formulation developed here.

Furthermore, as shown in Ref. [10], the expression for the virtual contribution is qualitatively consistent with the quasiclassical result obtained within the hyperspherical three-body formalism [15] when the interaction between the emitted protons is neglected.

Finally, adopting the quasiclassical approximation for deep sub-barrier one-proton decay widths [8], the relative magnitudes of the sequential and virtual contributions can be analyzed. As demonstrated in Ref. [10], sufficiently small values of the one-proton decay energy lead to a pronounced enhancement of the virtual contribution, which may exceed the sequential component.

## III. TOTAL AND PARTIAL WIDTHS OF DIAGONAL TWO-PROTON TWO-STEP VIRTUAL DECAYS

**OF SPHERICAL NUCLEI IN THE SUPERFLUID MODEL OF THE NUCLEUS**

To illustrate the formalism developed in the preceding section, we consider the diagonal two-step two-proton decay of a spherical parent nucleus $(Z, A)$ within the framework of the superfluid nuclear model [34] and the approach developed in Refs. [27–29]. The parent nucleus is assumed to decay from its ground state $J_A\,\sigma_A$ characterized by the quantum numbers and the corresponding intrinsic configuration. The decay is treated as two successive one-proton emissions.

During the first step, a proton occupying the shell-model orbital $nlj$ is emitted from the parent nucleus, resulting in the formation of an intermediate nucleus. In the second step, another proton is emitted from the same orbital of the intermediate nucleus, leading to the ground state of the final nucleus. Within this approximation, the total width of the diagonal two-proton decay is expressed as the sum of the corresponding partial contributions [27–29],

$$\Gamma = \sum_{J_{A-1}\sigma_{A-1}nlj} \Gamma(J_A\sigma_A \to J_{A-1}\sigma_{A-1}, nlj \to J_{A-2}\sigma_{A-2}) \qquad (40)$$

where the quantities $\Gamma(J_A\sigma_A \to J_{A-1}\sigma_{A-1}, nlj \to J_{A-2}\sigma_{A-2})$ denote the partial widths of the first and second one-proton transitions, respectively.

The partial widths depend on the kinetic energies of the emitted protons $T_1$ and $T_2 = \{Q_0(J_A\,\sigma_A; J_{A-2}\,\sigma_{A-2}) - T_1\}$, which are determined from the corresponding energy balance relations,

$$\Gamma(J_A\sigma_A \to J_{A-1}\sigma_{A-1}, nlj \to J_{A-2}\sigma_{A-2}) =$$
$$= \frac{1}{2\pi} \int_0^{Q_0(J_A\sigma_A;J_{A-2}\sigma_{A-2})} dT_1 \sum_{J_{A-1}\sigma_{A-1}} \left| \frac{\sqrt{\Gamma^{A}_{p_1}(J_A\sigma_A \to J_{A-1}\sigma_{A-1}, nlj)}\sqrt{\Gamma^{A-1}_{p_2}(J_{A-1}\sigma_{A-1} \to J_{A-2}\sigma_{A-2}, nlj)}}{Q_1(J_A\sigma_A; J_{A-1}\sigma_{A-1}) - T_1} \right|^2 \qquad (41)$$

The one-proton decay energies $Q_1(J_A\sigma_A;J_{A-1}\sigma_{A-1})$ and $Q_2(J_{A-1}\sigma_{A-1};J_{A-2}\sigma_{A-2})$ entering Eq. (41), together with the total two-proton decay energy $Q_0(J_A\sigma_A;J_{A-2}\sigma_{A-2})$, are expressed through the ground-state energies of the parent $E_0(J_A\sigma_A)$, intermediate Eq.(42-44), and final $E_0(J_{A-2}\sigma_{A-2})$ nuclei according to

$$Q_1(J_A\sigma_A;J_{A-1}\sigma_{A-1}) = E_0(J_A\sigma_A) - E(J_{A-1}\sigma_{A-1}), \tag{42}$$

$$Q_2(J_{A-1}\sigma_{A-1};J_{A-2}\sigma_{A-2}) = E(J_{A-1}\sigma_{A-1}) - E_0(J_{A-2}\sigma_{A-2}), \tag{43}$$

$$Q_0(J_A\sigma_A;J_{A-2}\sigma_{A-2}) = E_0(J_A\sigma_A) - E_0(J_{A-2}\sigma_{A-2}). \tag{44}$$

The amplitudes determining the deep sub-barrier one-proton decay widths are calculated using the integral formalism developed in Refs. [30–32],

$$\sqrt{\Gamma^A_{p_1}(J_A\sigma_A \to J_{A-1}\sigma_{A-1}, nlj)} = \sqrt{2\pi}\cdot G^A_{nlj}\int_0^\infty \tilde{F}_l(k_1,r)V_{pA-1}(r)R_{nlj}(r)dr \tag{45}$$

where $\tilde{F}_l(k_1,r)$is the energy-normalized radial Coulomb wave function of the emitted proton, $V_{pA-1}$ is the effective proton–nucleus interaction potential, and $q$ is the parameter used to adjust the depth of the single-particle potential in calculations of one- and two-proton decay characteristics.

The single-particle potential is taken in the Woods–Saxon form adopted in Refs. [35–37], which has previously been shown to provide a satisfactory description of spherical nuclei over a broad range of proton and neutron numbers and has been successfully applied to one-proton radioactivity in Refs. [31,32]:

$$V_N = \frac{-|V_0|}{1+\exp\left[\dfrac{r-R_0}{a}\right]}, \quad V_{ls} = -\frac{\chi}{r}\cdot\frac{dV_N(r)}{dr}(\mathbf{l}\cdot\mathbf{s}), \;\; R_0 = r_0 A^{1/3}, \tag{46}$$

The corresponding parameter values are listed in Table I.

The radial single-particle wave function of the proton occupying the shell-model state $nlj$ satisfies the Schrödinger equation

$$\left[-\frac{\hbar^2}{2m}\frac{d^2R_{nlj}(r)}{dr^2}+\frac{\hbar^2}{2m}\frac{l(l+1)}{r^2}+V_{pA-1}(r)+V_{pA-1}^{coul}(r)\right]R_{nlj}(r)=E(j)R_{nlj}(r), \quad (47)$$

Where $V_{pA-1}^{coul}(r)$ the Coulomb interaction between the proton and the daughter $(Z-1, A-1)$ nucleus is included explicitly.

The one-proton decay amplitudes also contain the proton spectroscopic (genealogical) factor $G_{nlj}^{A}$,

$$G_{nlj}^{A}(J_{P_A}\sigma_{P_A};J_{P_{A-1}}\sigma_{P_{A-1}},nlj)=\left\langle\left\{\psi_{\sigma_{P_{A-1}}}^{J_{P_{A-1}}m_{P_{A-1}}}a_{nljm}\right\}_{J_{P_A}m_{P_A}}\middle|\psi_{\sigma_{P_A}}^{J_{P_A}m_{P_A}}\right\rangle, \quad (48)$$

which determines the overlap between the parent $\psi_{\sigma_{P_A}}^{J_{P_A}m_{P_A}}$ and daughter $\psi_{\sigma_{P_{A-1}}}^{J_{P_{A-1}}m_{P_{A-1}}}$ nuclear wave functions. For even- and odd-$Z$ nuclei within the superfluid nuclear model, this quantity is given by [33]

$$\begin{cases} G_{nlj}^{A}=\sqrt{(2j+1)}v_j^i(-1)^l, \quad Z-\text{even}, \\ G_{nlj}^{A-1}=u_j^f, \quad Z-\text{odd}. \end{cases} \quad (49)$$

The nuclear structure calculations are performed within the superfluid model of atomic nuclei [33]. For even-$Z$ nuclei, the ground-state energy $E_{P_0}(J_A\sigma_A)$ and the corresponding Bogoliubov equations are written as

$$\frac{G}{2}\sum_j\frac{(j+\frac{1}{2})}{\varepsilon(j)}=1, \quad (50)$$

$$E_{P_0}(J_A\sigma_A)=\sum_j(2j+1)E(j)v_j^2-\frac{C_Z^2}{G}, \quad (51)$$

where

(52)

$$\varepsilon(j) \equiv \sqrt{C_Z^2 + \{E(j) - \lambda_Z\}^2}$$

The quantities $v_j, u_j$ denote the Bogoliubov transformation coefficients,

$$v_j^2 = \frac{1}{2}\left\{1 - \frac{E(j) - \lambda_Z}{\varepsilon(j)}\right\}, \; u_j^2 = 1 - v_j^2, \tag{53}$$

while the pairing interaction parameter $G$ is expressed as [34]

$$G = \frac{A_0}{A} \tag{54}$$

where coefficient $A_0$ = 16.5-17.5, and $C_Z$ and $\lambda_Z$ are the correlation function and chemistry potential of the proton subsystem.

For nuclei with an odd number of protons $Z$ (the total spin of an odd proton is equal to $j_2$), the blocking effect of the unpaired proton modifies the corresponding equations. In this case, the ground-state energy $E_P$ and the Bogoliubov equations take the form

$$\frac{G}{2}\left\{\sum_{j \neq j_2} \frac{j + 1/2}{\varepsilon(j, j_2)} + \frac{j - 1/2}{\varepsilon(j_2, j_2)}\right\} = 1 \tag{55}$$

$$Z = 1 + (2j_2 - 1)v_j^2(j_2) + \sum_{j \neq j_2} (2j + 1)v_j^2(j_2); \tag{56}$$

$$E_P = E(j_2) + (2j_2 - 1)E(j_2)v_j^2(j_2) + \sum_{j \neq j_2} (2j + 1)E(j)v_j^2(j_2) - \frac{C_Z^2(j_2)}{G}, \tag{57}$$

$$\varepsilon(j, j_2) \equiv \sqrt{C_Z^2(j_2) + \{E(j) - \lambda_Z(j_2)\}^2}, \tag{58}$$

$$v_j^2(j_2) = \frac{1}{2}\left\{1 - \frac{E(j) - \lambda_Z(j_2)}{\varepsilon(j, j_2)}\right\}; \; u_j^2(j_2) = 1 - v_j^2(j_2). \tag{59}$$

## IV. CALCULATION OF TOTAL AND PARCIAL WIDES OF DIAGONAL TWO-STEP TWO-PROTON DECAY OF $^{45}$FE NUCLEUS

The energies of one-proton decays (42) and (43) with proton escape and for the ground states of the parent and intermediate nucleus and 2p decay (44) can be found using the binding energies of the ground states of $^{45}Fe$, $^{44}Mn$, $^{43}Cr$ nuclei: $E_0(Z,A)$= (–329.308±0.078) MeV, $E_0(Z-1,A-1)$= (–329.284±0.076) MeV, $E_0(Z-2,A-2)$= (–330.462 ± 0.052) MeV. As a result, the energies (3-5) calculated using the isospin symmetry property of nuclei [38] are defined as $Q_1$ = (–0.024 ± 0.109) MeV, $Q_2$= (1.178 ± 0.92) MeV, and $Q_0$= (1.154 ± 0.142) MeV. The obtained energy value $Q_0$ is in the range of experimental values $Q_0^{\exp}$ = (1.154 ± 0.016) MeV [17-18], which will be used in further calculations. At the same time, the obtained value $Q_1$ = (–0.024 ± 0.109) MeV, $Q_2$= (1.178 ± 0.92) and $Q_0$= (1.154 ± 0.142). In principle, leads to the possibility of the appearance of real single-proton decay for positive values lying in the found interval 0 ≤ $Q_1$ ≤ 0.085 MэВ. Since the experimental energy distributions of emitted protons, in principle, do not indicate the possibility of the existence of this real one-proton decay, in further calculations for the value $Q_1$ will be used a negative value $Q_1$ = − 0.024 MeV, which weakly affects the calculated values of the total and partial widths of the 2p decay of the $^{45}Fe$ nucleus.

In calculations of the 2p decay widths, we fitted the energies $Q_1(J_A\sigma_A; J_{A-1}\sigma_{A-1})$ and $Q_0(J_A\sigma_A; J_{A-2}\sigma_{A-2})$ of both one-proton and 2p decays according to formulas (3-5) to their numerical values obtained above by varying the depth of the nuclear potential $V_{pA-1}(r)$ (through a change in the parameter $q$) and the superfluid parameter $G$. Table 2 shows the shell energies of the $^{45}Fe$ nucleus calculated for nuclear potentials [35-37] without fitting, and Table 3 shows the same energies at the parameter value $G$ = 0.375, and with fitting of the depth of the potential pit by varying the parameter $q$, the values of which are given in Table 3. The values $v_j^i$ of the parameters for the $^{45}Fe$ nucleus and $u_j^f$ for the $^{43}Cr$ nucleus are given in Tables 4-5. Table 6 gives the values of the overlap coefficients $G_{nlj}^A$ and $G_{nlj}^{A-1}$, as well as their products used in the calculations of the full and partial widths of the 2p decay of the $^{45}$Fe nucleus for the shell levels $2p_{3/2}$, $2p_{1/2}$, $2s_{1/2}$, $1d_{5/2}$, $1d_{3/2}$,$1f_{7/2}$, $1f_{5/2}$. The values of correlation functions and chemical potentials $\lambda_Z$ for $^{45}$Fe, $^{44}$Mn, and $^{43}$Cr nuclei given in Table 7 were within the range of variation of these values presented in [34].

The values of the total $\Gamma$ and partial widths of the 2p decay of the $^{45}$Fe nucleus calculated for the potentials [36-38] in the framework of the superfluid model of the atomic nucleus for the shell levels $2p_{3/2}$, $2p_{1/2}$, $2s_{1/2}$, $1d_{5/2}$, $1d_{3/2}$,$1f_{7/2}$, $1f_{5/2}$ and the values of the 2p decay energies $Q_0$ lying in the experimental range $Q_0$= (1.154 ± 0.016) MeV [18] are presented in Tables 8, 9. Table 9 shows that the widths of the levels $1d_{5/2}$, $1d_{3/2}$,$1f_{7/2}$, $1f_{5/2}$ are more than 3 orders of magnitude smaller than the experimental value of the total 2p decay width $\Gamma^{exp}$ = (1.6 ± 0.5)·10$^{-19}$ MeV, and therefore they can be disregarded in the future. Tables 8, 9 show a strong dependence of the calculated widths on the choice of shell potentials [35-37] and energy values $Q_0$. The changes in the partial widths $\Gamma\left(2p_{3/2}\right)$, $\Gamma\left(2p_{1/2}\right)$, $\Gamma\left(2s_{1/2}\right)$ and total width *Γ* are characterised by factors reaching values of 30, 27, 3, and 24 depending on the given choice. Table 10 presents the ratios of partial widths $\Gamma(2p_{3/2})/\Gamma(2s_{1/2})$, $\Gamma(2p_{3/2})/\Gamma(2p_{1/2})$, and the ratio $\Gamma/\Gamma^{exp}_{min}\,(\Gamma/\Gamma^{exp}_{max})$, where the values $\Gamma^{exp}_{min}$ = 1.1·10$^{-19}$ MeV and $\Gamma^{exp}_{max}$ = 2.1·10$^{-19}$ MeV correspond to the limits of the interval of the experimental values of the full width [18] given above.

Table 10 shows that the values of the calculated width G are close to the experimental values for the potentials [35-36], with the ratio $\Gamma/\Gamma^{exp}_{min}\,(\Gamma/\Gamma^{exp}_{max})$ being equal to 1.04 (0.54) at $Q_0$ = 1.170 MeV for the potential [4] and 1.20 (0.62) at $Q_0$ = 1.138 MeV for the potential [20] , and the ratio of the widths varies $\Gamma(2p_{3/2})/\Gamma(2p_{1/2})$ in the range 5-12. Therefore, for the calculations of the angular distributions of protons flown in the 2p decay of the $^{45}$Fe nucleus in the next section, it is sufficient to consider only the amplitudes of the partial widths of the 2p decay. The ratio is 9.2 for the potential [35] and 43.8 for the potential [36] at the above energy values.

It should be emphasized that the variation of the parameter q within the range of 1–

1.07 and the superfluid parameters λ and Δ within the limits indicated in Table 1. VII, does not go beyond the standard procedure adopted in such calculations (see, e.g., [31,32]). The values of λ and Δ for $^{45}$Fe, $^{44}$Mn and $^{43}$Cr are in the intervals characteristic of this region of nuclei and do not contradict the results of the work [33]. Thus, the fitting is not arbitrary, but is the matching of the microscopic parameters of the model with the known binding energies, which is a generally accepted method in nuclear physics. Moreover, the final agreement with the experimental width is achieved only with a well-defined combination of parameters, which indicates the predictive power of formalism, rather than fitting to specific data.

## V. ANGULAR DISTRIBUTION OF EMITTED PROTONS FOR TWO-PROTON DESTRUCTION OF A $^{45}$FE NUCLEUS

Using the two-step two-proton decay formalism developed in Refs. [27–29] together with the superfluid nuclear model [34], we derive the angular distribution $W(\mathbf{k}_1,\mathbf{k}_2)$ of protons emitted in the ground-state two-proton decay of $^{45}$Fe leading to the ground state of $^{43}$Cr. The distribution depends on the wave vectors $\mathbf{k}_1$ and $\mathbf{k}_2$ of the first and second emitted protons, whose directions in the laboratory frame are specified by the solid angles $\Omega_{\mathbf{k}_1}$ and $\Omega_{\mathbf{k}_2}$.

Assuming that the emitted proton pair forms a singlet state with total spin ($S$=0), which provides the dominant contribution to the decay, the normalized angular distribution is given by

$$W(\mathbf{k}_1,\mathbf{k}_2)=\frac{1}{\Gamma}\left|\sum_{lm}\sqrt{\frac{\Gamma(nlj_l)}{2l+1}}Y_{lm_l}(\Omega_{\mathbf{k}_1})Y_{l-m_l}(\Omega_{\mathbf{k}_2})\right|^2 \qquad (60)$$

where $\sqrt{\Gamma(nlj_l)}$ denotes the partial decay amplitude corresponding to the emission of the first proton with orbital angular momentum and the second proton with orbital angular momentum. The amplitudes associated with the dominant partial widths obtained in Sec. IV are retained in the calculation.

To simplify the analysis, we introduce a coordinate system in which the $z'$-axis is chosen along the momentum of the first emitted proton. In this frame, the spherical harmonics are transformed accordingly, while the polar angle of the first proton becomes zero. The direction of the second proton $\mathbf{k}_2$ is then completely determined by the opening angle $\Omega_{\mathbf{k}_2,\mathbf{k}_1}$ between the two emitted protons.

Using the transformation

$$Y_{lm_l}(0) = \sqrt{\frac{2l+1}{4\pi}}\delta_{m_l,0} \tag{61}$$

and integrating over the direction of the first proton, Eq. (60) reduces to

$$W(\theta) = \frac{1}{\Gamma}\left|\sum_l \sqrt{\Gamma(nlj_l)} Y_{l0}(\theta)\right|^2 \tag{62}$$

where $\theta = \theta_{\mathbf{k}_2,\mathbf{k}_1}$ is the angle between the directions of the vectors $\mathbf{k}_2$ and $\mathbf{k}_1$. From (62), the normalized angular distribution is obtained as

$$\bar{W}(\theta) = \frac{2\pi}{\Gamma}\sum_{ll'} \sqrt{\Gamma(nlj_l)}\sqrt{\Gamma(nl'j_{l'})} Y_{l0}(\theta) Y_{l'0}(\theta)\sin\theta \tag{63}$$

subject to the normalization condition

$$\int_0^\pi \bar{W}(\theta)\, d\theta = 1 \tag{64}$$

The representation (63) is particularly convenient because it can be compared directly with the experimentally measured unit-normalized angular distributions $\bar{W}^{\exp}(\theta)$ reported in Refs. [20,21, 40]. The corresponding experimental distribution is shown in Fig. 3.

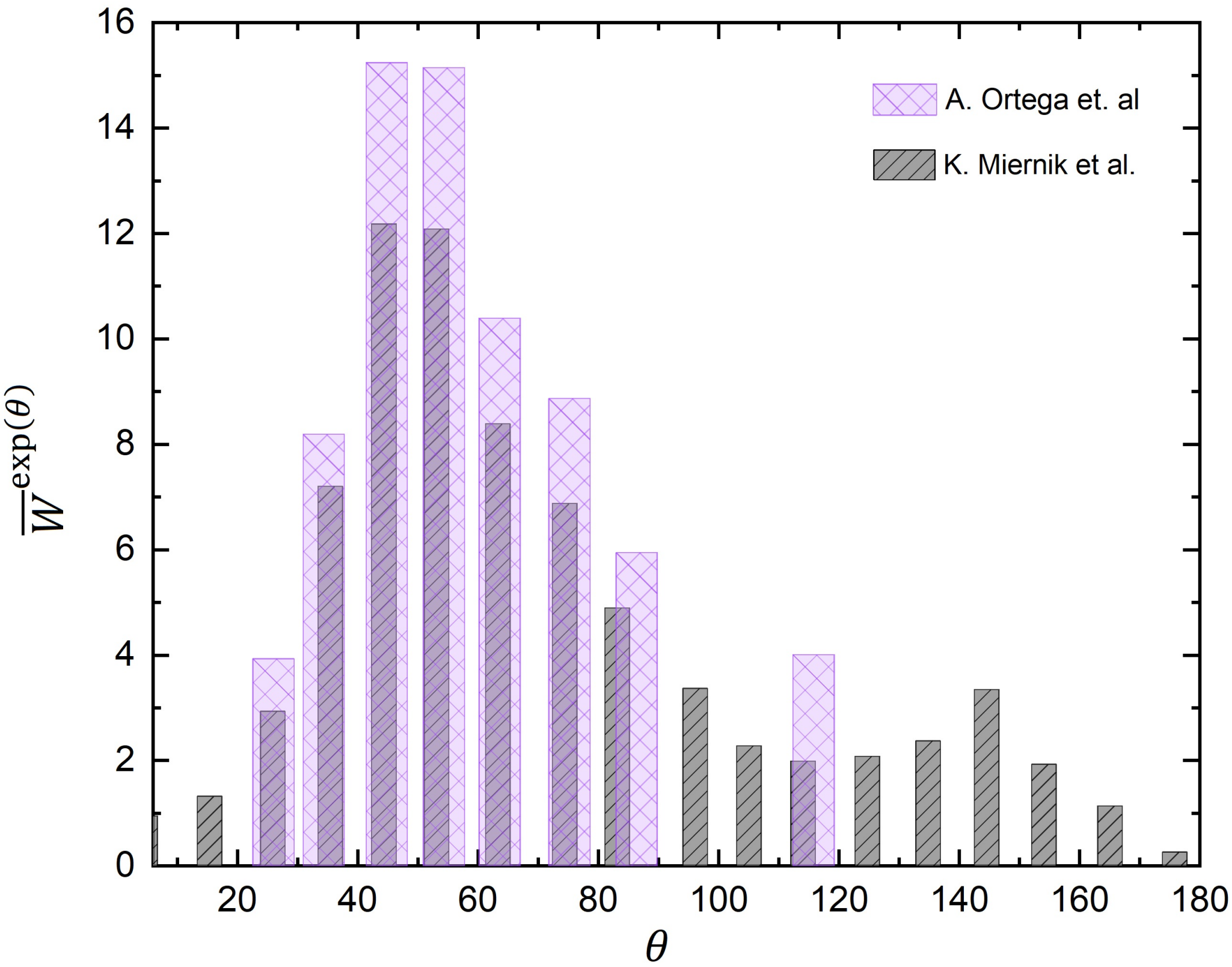

FIG.3. Unit-normalised experimental [21, 41] angular distribution $\overline{W}^{\exp}(\theta)$ of protons - products of the 2p decay of the $^{45}$Fe nucleus, presented as a histogram.

The symmetry properties of the spherical harmonics $Y_{l0}(\theta)$,

$$Y_{l0}(\theta) = (-1)^l\, Y_{l0}(\pi - \theta) \tag{65}$$

show that a symmetric angular distribution is expected when only partial waves of identical parity contribute. Consequently, the experimentally observed asymmetry of the proton angular distribution can only arise from the interference between decay amplitudes corresponding to orbital angular momenta of opposite parity.

Retaining only the dominant partial widths $\Gamma(2p_{3/2})$ and $\Gamma(2s_{1/2})$ identified in the previous section and introducing the ratio $\beta^2 = \Gamma(2p_{3/2})/\Gamma(2s_{1/2})$, Eq. (63) can be reduced to the one-parameter expression

$$\overline{W}(\theta) = \frac{1}{2(1+\beta^2)}\left(1+\sqrt{3}\beta\cos\theta\right)^2 \sin\theta \tag{66}$$

To analyse the possibility of describing by the one-parameter expression (66) the experimental angular distribution of protons $\overline{W}^{\exp}(\theta)$ [20-21], we use the fitting method [39]. The calculations show that the minimal deviations $\overline{W}^{\exp}(\theta)$ from $\bar{W}(\theta)$= are $\beta$ = $\beta_0$ = 0.21; 1.7; 3.

A comparison of the calculated values $\beta = \sqrt{\Gamma(2p_{3/2})/\Gamma(2s_{1/2})}$ (Table 10) and the value $\beta_0$ = 0.21 leads to the conclusion that in the intervals of experimental values of both the 2p decay energy $Q_0$= (1.154 ± 0.016) MeV and the total decay width $\Gamma^{\exp}$= (1.6 ± 0.5)·$10^{-19}$ MeV for all three sets of parameters of the considered shell potentials [36-38], the indicated value $\beta_0$ does not agree with the calculated values $\beta$. The agreement of $\beta_0$= 1.7 with analogue values of the quantities is present only for the potential [38]. But for this potential the calculated value of the total width Γ differs from its experimental value by a factor greater than 4, so that $\beta_0$= 1.7 can also be excluded from consideration. Finally, $\beta_0$= 3 reasonably agrees with the values of $\beta$ (Table 10) only for the potential [35] at the upper limit of the 2p decay energy = 1.17 MeV and the lower limit of the experimental width $\Gamma^{\exp}_{\min}$ = 1.1·$10^{-19}$ MeV. The comparison of the normalised angular distribution $\overline{W}(\theta)$ at $\beta_0$ = 3 and the experimental $\overline{W}^{\exp}(\theta)$, shown in FIG.4, demonstrates their qualitative agreement.

To enable a meaningful comparison between the unit-normalized theoretical probability density $\overline{W}(\theta)$ and the experimental data, the calculated curves were multiplied by an overall scaling factor. This factor accounts for the total number of recorded experimental events and the angular bin width of $10^{\circ}$. As a result of this standard normalization procedure, the theoretical curve for $\beta_0 = 3$ not only automatically reproduces the experimental amplitude of 14–16 counts in the primary peak region ($\theta = 40^{\circ}$), but also accurately describes the overall shape of the entire angular distribution.

Finally, the combined analysis of the total decay width presented in Sec. IV and the angular distribution considered here indicates that only the shell-model potential of Ref. [36] provides a simultaneous description of both observables within the experimental uncertainties. This result demonstrates the strong sensitivity of two-proton radioactivity to the underlying single-particle potential and confirms that the angular correlation of the emitted protons provides an additional and independent constraint on the microscopic description of the decay mechanism.

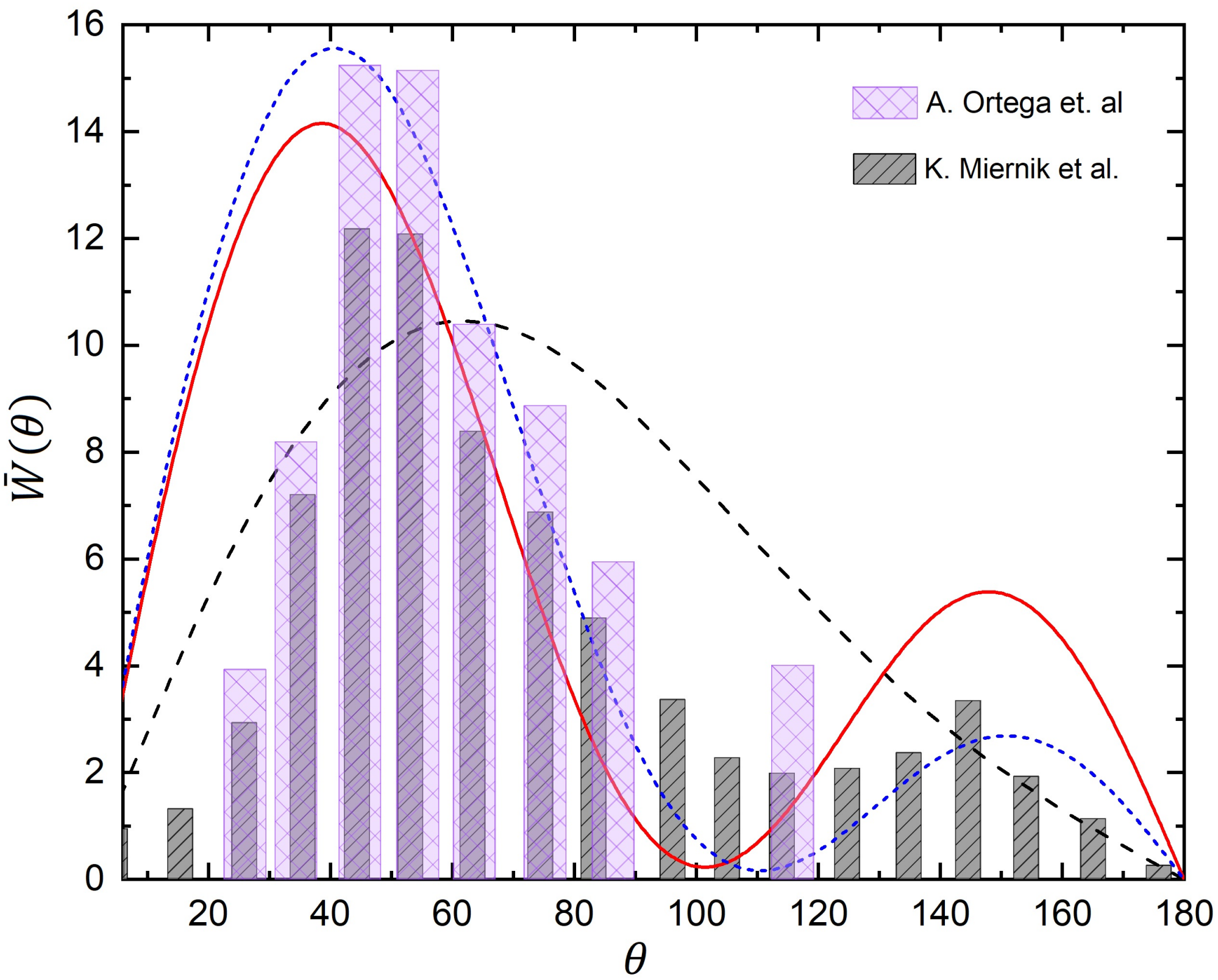

FIG. 4. Comparison of the experimental [21, 40] $\overline{W}^{\exp}(\theta)$ (histogram) and calculated by formulas (25) $\overline{W}(\theta)$ at different values of the phytised parameter $\beta_0$: dashed-dotted curve - $\beta_0 = 0.21$; dotted curve - $\beta_0 = 1.7$; solid curve is $\beta_0 = 3$.

## VI. CONCLUSION

In the present work, a general quantum-mechanical theory of two-step (taking into account strictly the effects of pairing and superfluid correlations of nucleons in the parent $(Z,A)$, intermediate $(Z-1,A-1)$, and final $(Z-2,A-2)$ nuclei). two-proton decay has been developed on the basis of the multiparticle formalism of one-proton radioactivity. The proposed approach describes two-proton emission as two successive one-proton transitions connected by the Green's function of the intermediate nucleus and provides a unified treatment of both sequential and virtual decay mechanisms within a single theoretical framework.

Unlike conventional quasiclassical approaches, the present formalism is derived entirely within quantum mechanics and does not require the introduction of channel radii. Since the theory is formulated in terms of microscopic one-proton decay amplitudes, it consistently incorporates the nuclear structure of the initial, intermediate, and final nuclei, including shell structure, angular-momentum coupling, pairing correlations, and superfluid effects. Consequently, the formalism can be applied to both spherical and deformed nuclei as well as to diagonal and, for spherical nuclei, off-diagonal one-proton transitions.

The derived expressions for the sequential and virtual two-proton decay widths naturally arise from the same Green-function formalism. The sequential contribution reduces to the well-established results of the R -matrix theory [11] and the kinetic-equation approach [12], providing an important consistency check of the present formulation. At the same time, the virtual contribution extends the theoretical description to off-shell intermediate configurations within the same microscopic framework.

The developed formalism has been applied to the ground-state two-proton decay of $^{45}Fe$ leading to the ground state of $^{43}Cr$. Numerical calculations performed within the superfluid nuclear model demonstrate a pronounced dependence of the calculated total and partial decay widths on the choice of the single-particle shell potential. Among the potentials considered in the present work, the parameterization of Ref. [35] provides the only simultaneous description of both the experimental total decay width and the measured proton angular distribution.

This is justified by the fact that in the energy range under consideration (about 1 MeV) and for nuclei with a sufficiently large charge ($Z = 26$), the Coulomb repulsion between protons makes a correction of the order of several percent to the full width, which is significantly less than the observed sensitivity to the choice of a single-particle potential (which reaches a factor of 10–20). In the serial channel, the interaction between protons is suppressed due to the separation in departure time. For a virtual channel, the inclusion of the proton-proton Coulomb interaction would require a modification of the three-particle Green function, which is beyond the scope of this work, but can be considered in the future as a development of formalism. Nevertheless, the good agreement with the experiment suggests that neglecting this interaction is a permissible first approximation.

**ACKNOWLEDGEMENTS**

The authors express their gratitude to the blessed memory of Yuri Vladimirovich Ivan kov, Associate Professor of the Department of Nuclear Physics, whose many years of contribution to the study of two-proton decay of nuclei have become the foundation fo r the development of this research direction. The scientific ideas and methodological a pproaches of Yu.V. Ivankov remain relevant and continue to serve as a guiding refere nce for researchers in the field of nuclear physics.

This work was supported by the Russian Science Foundation (Project No. 25-22-0069 7).

1. P. I. Woods and C. N. Davids, *Annu. Rev. Nucl. Part. Sci.* **47**, 54 (1997).

2. S. G. Kadmensky and V. E. Kalechits, *Yad. Fiz.* **12**, 70 (1970).

3. S. G. Kadmensky and V. I. Furman, *Alpha Decay and Related Nuclear Reactions* (Energoatomizdat, Moscow, 1985).

4. S. G. Kadmensky and V. G. Khlebostroev, *Yad. Fiz.* **18**, 980 (1973).

5. V. P. Bugrov, S. G. Kadmensky, and V. I. Furman, *Yad. Fiz.* **41**, 1123 (1985).

6. V. P. Bugrov, V. E. Bunakov, S. G. Kadmensky, and V. I. Furman, *Yad. Fiz.* **42**, 57 (1985).

7. V. P. Bugrov and S. G. Kadmensky, *Yad. Fiz.* **49**, 1562 (1989).

8. V. I. Goldansky, *Nucl. Phys.* **19**, 482 (1960).

9. L. V. Grigorenko, *Phys. Part. Nucl.* **40**, 1277 (2009).

10. S. G. Kadmensky and Yu. V. Ivankov, *Yad. Fiz.*, in press.

11. A. M. Lane and A. G. Thomas, *Rev. Mod. Phys.* **30**, 257 (1958).

12. E. Segre, *Experimental Nuclear Physics*, Vol. 3 (IIL, Moscow, 1961), p. 13.

13. L. V. Grigorenko *et al*., *Phys. Rev. Lett.* **85**, 22 (2000).

14. L. V. Grigorenko *et al*., *Phys. Rev. C* **64**, 054002 (2001).

15. L. V. Grigorenko and M. V. Zhukov, *Phys. Rev. C* **76**, 014009 (2007).

16. V. I. Goldansky, *Nucl. Phys.* **19**, 482 (1960).

17. M. Pfutzner *et al*., *Eur. Phys. J. A* **14**, 279 (2002).

18. J. Giovanezzo *et al*., *Phys. Rev. Lett.* **89**, 102501 (2002).

19. C. Dossat *et al*., *Phys. Rev. C* **72**, 054315 (2005).

20. K. Miernik *et al*., *Phys. Rev. Lett.* **72**, 054315 (2005).

21. K. Miernik *et al*., *Nucl. Instrum. Methods Phys. Res. A* **581**, 194 (2007).

22. B. A. Brown and F. C. Barker, *Phys. Rev. C* **67**, 041309 (2003).

23. L. D. Landau and Ya. A. Smorodinsky, *Zh. Eksp. Teor. Fiz.* **14**, 269 (1944).

24. A. B. Migdal, *Zh. Eksp. Teor. Fiz.* **28**, 3 (1955).

25. K. M. Watson, *Phys. Rev.* **88**, 1163 (1952).

26. L. V. Grigorenko, *Phys. Part. Nucl.* **40**, 1273 (2009).

27. S. G. Kadmensky and Yu. V. Ivankov, *Yad. Fiz.* **77**, 1075 (2014) [*Phys. Atom. Nucl.* **77**, 1075 (2014)].

28. S. G. Kadmensky and Yu. V. Ivankov, *Yad. Fiz.* **77**, 1605 (2014) [*Phys. Atom. Nucl.* **77**, 1605 (2014)].

29. S. G. Kadmensky and Yu. V. Ivankov, *Izv. RAN, Ser. Fiz.* **78**, 1414 (2014) [*Bull. Russ. Acad. Sci. Phys.* **78**, 1414 (2014)].

30. S. G. Kadmensky and V. I. Furman, *Alpha Decay and Related Nuclear Reactions* (Energoatomizdat, Moscow, 1985).

31. V. P. Bugrov, S. G. Kadmensky, V. I. Furman, and V. G. Khlebostroev, *Yad. Fiz.* **41**, 1123 (1985) [*Sov. J. Nucl. Phys.* **41**, 717 (1985)].

32. V. P. Bugrov and S. G. Kadmensky, *Yad. Fiz.* **49**, 1562 (1989).

33. J. Rotureau, J. Okołowicz, M. Płoszajczak, Theory of the two-proton radioactivity in the continuum shell model, *Nuclear Physics A* **767**, 13–57 (2006)

34. V. G. Soloviev, *Theory of Atomic Nucleus: Nuclear Models* (Energoatomizdat, Moscow, 1981).

35. A. B. Migdal, *Theory of Finite Fermi Systems and Properties of Atomic Nuclei* (Nauka, Moscow, 1965).

36. S. A. Fayans, Preprint IAE-1593 (1968).

37. F. D. Becchetti and G. W. Greenless, *Phys. Rev.* **182**, 1190 (1969).

38. P. E. Nemirovsky, *Modern Models of the Atomic Nucleus* (Atomizdat, Moscow, 1980).

39. B. A. Brown, *Phys. Rev. C* **43**, 43 (1990).

40. G. Cramer, *Mathematical Methods of Statistics* (Mir, Moscow, 1975).

40. Ortega Moral A., Wang S. M., Giovinazzo J. и др. Two-proton correlations in the decay of $^{48}Ni$ and $^{45}Fe$ // Physical Review C. — 2025. — Т. 112. — С. L061302. — DOI: 10.1103/kqb5-z2xy.

## APPENDICES

**TABLE I.** Parameters of nuclear potentials.

| Potential | $V_0$, MeV | $r_0$, Fm | $\chi$, Fm$^2$ | $a$, Fm |
|---|---|---|---|---|
| [35] | 51.9+29.0$(N-Z)/A$ | 1.28 | 0.175 | 0.67 |
| [36] | 54.0+24.0$(N-Z)/A$ | 1.17 | 12.4/$V_0$ | 0.75 |
| [37] | 53.0+33.4$(N-Z)/A$ | 1.24 | 0.26·[1+2$(N-Z)/A$] | 0.63 |

**TABLE II.** Energies (MeV) of proton shell levels of the $^{45}$Fe nucleus calculated with the initial potentials [35-37] at q=1. with initial potentials [35-37] at q=1.

| *nlj* | [19] | [20] | [21] |
|---|---|---|---|
| $1f_{5/2}$ | 3.5554 | 5.9963 | 5.7808 |
| $2p_{1/2}$ | 3.3869 | 4.0703 | 4.6974 |
| $2p_{3/2}$ | 2.9878 | 3.5070 | 3.8517 |
| $1f_{7/2}$ | 1.8644 | 3.4647 | 2.3453 |
| $2s_{1/2}$ | –4.8655 | –4.1695 | –3.8399 |
| $1d_{3/2}$ | –6.8646 | –4.8268 | –5.2431 |
| $1d_{5/2}$ | –8.0278 | –7.1723 | –7.6657 |
| $1p_{1/2}$ | –16.9629 | –16.2860 | –16.1123 |
| $1p_{3/2}$ | –17.5382 | –17.7372 | –17.3060 |
| $1s_{1/2}$ | –25.7316 | –27.0819 | –25.6382 |

**TABLE III.** Same as in Table 2, $q > 1$.

| $nlj$ | [35] $q$ = 1.0058 | [36] $q$ = 1.0715 | [37] $q$ = 1.0195 |
|---|---|---|---|
| $1f_{5/2}$ | 3.2143 | 4.2203 | 5.0746 |
| $2p_{1/2}$ | 3.0037 | 2.6535 | 3.9827 |
| $2p_{3/2}$ | 2.5830 | 2.0062 | 3.1514 |
| $1f_{7/2}$ | 1.4258 | 1.4249 | 1.4784 |
| $2s_{1/2}$ | –5.5075 | –6.6747 | –4.9833 |
| $1d_{3/2}$ | –7.5201 | –7.5314 | –6.4631 |
| $1d_{5/2}$ | –8.7228 | –9.9422 | –8.9261 |
| $1p_{1/2}$ | –17.7913 | –19.5716 | –17.5984 |
| $1p_{3/2}$ | –18.3805 | –21.0419 | –18.7802 |
| $1s_{1/2}$ | –26.7873 | –30.8798 | –27.4175 |

**TABLE IV.** Coefficients for the $^{45}$Fe core.

| Potential | [35] | [36] | [37] |
|---|---|---|---|
| $nlj$ | $v_j$ | $v_j$ | $v_j$ |
| $1f_{5/2}$ | 0.293 | 0.276 | 0.219 |
| $2p_{1/2}$ | 0.397 | 0.435 | 0.288 |
| $2p_{3/2}$ | 0.461 | 0.555 | 0.394 |
| $1f_{7/2}$ | 0.711 | 0.710 | 0.623 |
| $2s_{1/2}$ | 0.991 | 0.994 | 0.993 |
| $1d_{3/2}$ | 0.995 | 0.995 | 0.995 |

| $1d_{5/2}$ | 0.996 | 0.997 | 0.997 |
|---|---|---|---|

**TABLE V.** Coefficients for the [43]Cr core.

| По-тен-циал | [35] | [36] | [37] |
|---|---|---|---|
| *nlj* | $u_j$ | $u_j$ | $u_j$ |
| $1f_{5/2}$ | 0.980 | 0.980 | 0.979 |
| $2p_{1/2}$ | 0.961 | 0.961 | 0.960 |
| $2p_{3/2}$ | 0.933 | 0.932 | 0.931 |
| $1f_{7/2}$ | 0.774 | 0.773 | 0.773 |
| $2s_{1/2}$ | 0.125 | 0.125 | 0.125 |
| $1d_{3/2}$ | 0.102 | 0.102 | 0.102 |
| $1d_{5/2}$ | 0.079 | 0.079 | 0.076 |

**TABLE VI.** Overlap coefficients $G_{nlj}^{A}, G_{nlj}^{A-1}$ and $G_{nlj}^{A} \cdot G_{nlj}^{A-1}$ calculated for potentials [19-21].

| Po-ten-tial | [35] | | | [36] | | | [37] | | |
|---|---|---|---|---|---|---|---|---|---|
| *nlj* | $G_{nlj}^{A}$ | $G_{nlj}^{A-1}$ | $G_{nlj}^{A} \cdot G_{nlj}^{A-1}$ | $G_{nlj}^{A}$ | $G_{nlj}^{A-1}$ | $G_{nlj}^{A} \cdot G_{nlj}^{A-1}$ | $G_{nlj}^{A}$ | $G_{nlj}^{A-1}$ | $G_{nlj}^{A} \cdot G_{nlj}^{A-1}$ |
| $1f_{5/2}$ | 0.718 | 0.980 | 0.704 | 0.676 | 0.980 | 0.662 | 0.536 | 0.979 | 0.525 |
| $2p_{1/2}$ | 0.561 | 0.961 | 0.539 | 0.615 | 0.961 | 0.591 | 0.407 | 0.960 | 0.391 |
| $2p_{3/2}$ | 0.922 | 0.933 | 0.860 | 1.110 | 0.932 | 1.035 | 0.788 | 0.931 | 0.734 |
| $1f_{7/2}$ | 2.011 | 0.774 | 1.557 | 2.008 | 0.773 | 1.552 | 1.762 | 0.773 | 1.362 |
| $2s_{1/2}$ | 1.401 | 0.125 | 0.175 | 1.406 | 0.125 | 0.176 | 1.410 | 0.125 | 0.176 |

| $1d_{3/2}$ | 1.990 | 0.102 | 0.203 | 1.990 | 0.102 | 0.203 | 1.990 | 0.102 | 0.203 |
|---|---|---|---|---|---|---|---|---|---|
| $1d_{5/2}$ | 2.440 | 0.079 | 0.193 | 2.442 | 0.079 | 0.193 | 2.442 | 0.076 | 0.186 |

**TABLE VII.** Values $C_Z, \lambda_Z$ for $^{45}$Fe, $^{44}$Mn and $^{43}$Cr.

| Nu-cleus | Potential | $C_Z$, MeV | $\lambda_Z$, MeV |
|---|---|---|---|
| $^{45}$Fe | [35] | 1.850 | 1.200 |
| | [36] | 1.755 | 1.194 |
| | [37] | 1.546 | 1.591 |
| $^{44}$Mn | [35] | 1.608 | 1.522 |
| | [36] | 1.747 | 1.113 |
| | [37] | 1.608 | 1.522 |
| $^{43}$Cr | [35] | 1.682 | 1.026 |
| | [36] | 1.698 | 1.005 |
| | [37] | 1.610 | 1.438 |

**TABLE VII.** Partial and total 2p decay widths of the $^{45}$Fe nucleus.

| Potential | $Q_0$, MэB | $\Gamma(2s_{1/2})\,10^{19}$, MeV | $\Gamma(2p_{3/2})\,10^{19}$, MeV | $\Gamma(2p_{1/2})\,10^{19}$, MeV | $\Gamma\,10^{19}$, MeV |
|---|---|---|---|---|---|
| [35] | 1.138 | 0.041 | 0.381 | 0.072 | 0.495 |
| | 1.154 | 0.055 | 0.505 | 0.096 | 0.657 |
| | 1.170 | 0.096 | 0.881 | 0.169 | 1.148 |
| [36] | 1.138 | 0.027 | 1.182 | 0.128 | 1.337 |
| | 1.154 | 0.036 | 1.566 | 0.170 | 1.772 |
| | 1.170 | 0.063 | 2.720 | 0.296 | 3.080 |
| [37] | 1.138 | 0.031 | 0.093 | 0.008 | 0.131 |
| | 1.154 | 0.041 | 0.123 | 0.011 | 0.175 |

|  | 1.170 | 0.071 | 0.215 | 0.019 | 0.306 |
|---|---|---|---|---|---|

**TABLE IX.** Partial widths of the 2p decay of the 45Fe nucleus.

| Potential | $Q_0$, MeV | $\Gamma(1d_{5/2})$, MeV | $\Gamma(1d_{3/2})$, MeV | $\Gamma(1f_{7/2})$, MeV | $\Gamma(1f_{5/2})$, MeV |
|---|---|---|---|---|---|
| [35] | 1.138 | $0.987\cdot10^{-24}$ | $0.109\cdot10^{-23}$ | $0.240\cdot10^{-22}$ | $0.362\cdot10^{-24}$ |
| | 1.154 | $0.132\cdot10^{-23}$ | $0.146\cdot10^{-23}$ | $0.318\cdot10^{-22}$ | $0.486\cdot10^{-24}$ |
| | 1.170 | $0.233\cdot10^{-23}$ | $0.258\cdot10^{-23}$ | $0.550\cdot10^{-22}$ | $0.862\cdot10^{-24}$ |
| [36] | 1.138 | $0.649\cdot10^{-24}$ | $0.681\cdot10^{-24}$ | $0.144\cdot10^{-22}$ | $0.690\cdot10^{-25}$ |
| | 1.154 | $0.869\cdot10^{-24}$ | $0.911\cdot10^{-24}$ | $0.190\cdot10^{-22}$ | $0.925\cdot10^{-25}$ |
| | 1.170 | $0.154\cdot10^{-23}$ | $0.161\cdot10^{-23}$ | $0.329\cdot10^{-22}$ | $0.164\cdot10^{-24}$ |
| [37] | 1.138 | $0.639\cdot10^{-24}$ | $0.706\cdot10^{-24}$ | $0.187\cdot10^{-22}$ | $0.169\cdot10^{-25}$ |
| | 1.154 | $0.855\cdot10^{-24}$ | $0.944\cdot10^{-24}$ | $0.248\cdot10^{-22}$ | $0.227\cdot10^{-25}$ |
| | 1.170 | $0.151\cdot10^{-23}$ | $0.166\cdot10^{-23}$ | $0.430\cdot10^{-22}$ | $0.405\cdot10^{-25}$ |

**TABLE X.** Values of the ratios $\Gamma(2p_{3/2})/\Gamma(2s_{1/2})$ , $\Gamma(2p_{3/2})/\Gamma(2p_{1/2})$, $\Gamma/\Gamma^{\exp}_{\min}$ ( $\Gamma/\Gamma^{\exp}_{\max}$ ) for the 2p decay of the $^{45}$Fe nucleus.

| Potential | $Q_0$, MeV | $\Gamma(2p_{3/2})/\Gamma(2s_{1/2})$ | $\Gamma(2p_{3/2})/\Gamma(2p_{1/2})$ | $\Gamma/\Gamma^{\exp}_{\min}$ | $\Gamma/\Gamma^{\exp}_{\max}$ |
|---|---|---|---|---|---|
| [35] | 1.138 | 9.29 | 5.29 | 0.45 | 0.23 |
| | 1.154 | 9.18 | 5.26 | 0.60 | 0.31 |
| | 1.170 | 9.18 | 5.21 | 1.04 | 0.54 |
| [36] | 1.138 | 43.78 | 9.23 | 1.20 | 0.62 |
| | 1.154 | 43.50 | 9.21 | 1.61 | 0.84 |
| | 1.170 | 43.17 | 9.19 | 2.8 | 1.46 |
| [37] | 1.138 | 3.00 | 11.63 | 0.12 | 0.06 |
| | 1.154 | 3.00 | 11.18 | 0.16 | 0.08 |
| | 1.170 | 3.03 | 11.32 | 0.28 | 0.14 |